\documentclass[aps,preprint]{revtex4}%
\usepackage{amsfonts}
\usepackage{amsmath}
\usepackage{amssymb}
\usepackage{graphicx}%
\begin{document}
\title{Triple-gap superconductivity of MgB$_{2}$ - (La,Sr)MnO$_{3}$ composite.\\Which of the gaps is proximity induced?}
\author{V. N. Krivoruchko and V. Yu. Tarenkov}
\affiliation{Donetsk Physics \& Technology Institute NAS of Ukraine, Str. R. Luxemburg 72,
83114 Donetsk, Ukraine}
\date{\today}


\begin{abstract}
Interplay of superconductivity and magnetism in a composite prepared of the
ferromagnetic half-metallic La$_{0.67}$Sr$_{0.33}$MnO$_{3}$ (LSMO)
nanoparticles and the conventional s-wave superconductor MgB$_{2}$ has been
studied. A few principal effects have been found in bulk samples. With an
onset of the MgB$_{2}$ superconductivity, a spectacular drop of the sample
resistance has been detected and superconductivity has been observed at
temperature up to 20K. Point-contact (PC) spectroscopy has been used to
measure directly the superconducting energy coupling. For small voltage, an
excess current and doubling of the PC's normal state conductance have been
found. Conductance peaks corresponding to three energy gaps are clearly
observed. Two of these gaps we identified as enhanced $\Delta_{\pi}$ and
$\Delta_{\sigma}$ gaps originating from the MgB$_{2}$; the third gap
$\Delta_{tr}$ is more than three times larger than the largest MgB$_{2}$ gap.
The experimental results provide unambiguous evidences for a new type of
proximity effect which follows the phase coherency scenario of proximity
induced superconductivity. Specifically, at low temperature, the
\textit{p}-wave spin-triplet condensate with pairing energy $\Delta_{tr}$ is
essentially sustained in LSMO but is incapable to display long-range
supercurrent response because of a phase-disordering state. The proximity
coupling to MgB$_{2}$ restores the long-range phase coherency of the triplet
superconducting state, which, in turn, enhances superconducting state of the
MgB$_{2}$.

\end{abstract}
\maketitle

When an itinerant ferromagnet (F) is in contact with an s-wave superconductor
(S), superconductivity is expected to decay rapidly (a few nanometers) inside
the F owing to the incompatible nature of superconductivity and ferromagnetic
order \cite{Bu05}. This expectation was indeed confirmed in various materials
and geometries. On the other hand, a series of recent experiments \cite{Gi98,
Pet99, Pe04, Sos06, Kei06, Wang10} present clear evidences that a simple
physical interpretation of the proximity effect, reading that the Cooper pairs
are broken by a strong exchange field in the F layer, is in reality too
simplistic, and an extension of the existing concepts of interplay between
superconductivity and ferromagnetism is needed. 

From the theoretical viewpoint, a case of an F with a uniform exchange field
is well understood and the proximity effect may be described by taking into
account the splitting of electronic bands of opposite spins \cite{Bu05}. The
situation becomes more complicated if the magnetic structure is inhomogeneous,
and only few works (see, e.g. \cite{Iv09}) have been devoted to the problem of
superconducting proximity systems with non-collinear magnetic disorder
correlated on length scales of an order or larger than the superconducting
coherence length. On the other hand, being an experimentally accessible
electronic system with tunable parameters, hetero-structures with
inhomogeneous magnetism (e.g., superconducting nano-composites) offer a unique
testing ground for studying superconducting proximity systems with an
arbitrary length of magnetic disorder. However, to our knowledge, up to now no
reports have been devoted to a superconductive composite of (nano)granular
half-metal ferromagnet (hmF) and a conventional s-wave superconductor.

In our recent works anomalous superconductivity has been detected in point
contacts (PCs) of half-metallic manganite, (La,Sr)MnO$_{3}$ and
(La,Ca)MnO$_{3}$, with an \textit{s}-wave superconductor, Pb or MgB$_{2}$,
\cite{Kr06, Kr07}. Particularly, for proximity affected PCs, the coherent
multiple Andreev reflection (sub-harmonic gap resonances) has been observed,
and it has been found that proximity induced superconducting gap of
(La,Sr)MnO$_{3}$ or (La,Ca)MnO$_{3}$ is much larger than that of the Pb or
MgB$_{2}$. It was implied that, at low temperature, \textit{non-coherent
superconducting fluctuations are essentially sustained in half-metallic
manganites and, in proximity affected region, the singlet S establishes phase
coherence of the p-wave spin-triplet superconducting state of the manganites
}\cite{Kr06, Kr07}. To verify this phase-disordering scenario for anomalous
superconductivity of proximity affected PCs, we prepared and systematically
studied normal and superconducting properties of the MgB$_{2}$ -- (nano)
La$_{0.67}$Sr$_{0.33}$MnO$_{3}$ (MgB:LSMO) composite. The basic idea was to
obtain such a composite, where proximity coupled hmF/S interfaces govern
superconducting properties of the sample. Fortunately, the idea was
successful, and in this report we present the results of investigating
transport properties of such systems. Our key findings are: (i) the samples of
MgB:LSMO (nano)composite demonstrate evidences for bulk (triplet)
superconductivity with large, up to 20K, critical temperature, T$_{C}$; (ii)
three superconducting gaps are clearly observed. We identified two of these
gaps as enhanced MgB$_{2}$ gaps, the third quasiparticle gap, $\Delta_{tr}$,
is more than three times larger than the largest MgB$_{2}$ gap and could not
be attributed to this superconductor. Also, we found that (iii) all the energy
gaps vanish simultaneously as the temperature increases towards T$_{C}$ of
MgB$_{2}$.

The origin of the large quasiparticle $\Delta_{tr}$ gap and an enhancement of
the MgB$_{2}$ gaps cannot be explained on the basis of the existing
theoretical models. We believe the results obtained are the first observation
of the new type of superconducting proximity effect. That is, MgB$_{2}$
enhances the internal superconducting order parameter phase stiffness in LSMO
and, in turn, being in superconducting state, LSMO enhances superconducting
characteristics of MgB$_{2}$. Taking into account half-metallic properties of
LSMO and the Curie temperature magnitude, we think that in the case of LSMO we
deal with even frequency \textit{p}-wave spin-triplet superconducting pairing.

\textit{Experimental details}. Stoichiometric MgB$_{2}$ has a transition
temperature T$_{C}$ = 39 K, the highest among conventional superconductors
\cite{XX}. MgB$_{2}$, however, is by no means an ordinary, but the weakly
interacting multiple bands superconductor ($\sigma$ and $\pi$ bands). It
demonstrates distinct multiple superconducting energy gaps (the $\sigma$ and
$\pi$ gaps) with energy $\Delta_{\pi}$ = 2.3 meV and $\Delta_{\sigma}$ = 7.1
meV at T = 4.2K \cite{XX}. The coherence length $\xi$ is anisotropic with
$\xi_{ab}$(0) $\approx$\ 8nm and $\xi_{c}$(0) $\approx$\ 2nm \cite{XX}.
La$_{0.67}$Sr$_{0.33}$MnO$_{3}$ is a hmF with the T$_{Curie}$ = 320K (see,
e.g., \cite{man}). Composites containing submicron MgB$_{2}$ powder and LSMO
particles (about 20nm in size) have been produced. Details of the LSMO
nanoparticles preparation may be found in Ref. \cite{Sav}. The powders with
different weight ratios of LSMO and MgB$_{2}$ were mixed and pressed (under
pressure up to 60 kbar) in stripes, and a number of samples of different
composition have been obtained.

The obtained composites were analyzed by the method of X-ray energy-dispersive
spectroscopy (EDX) using INCAPentaFETx3 spectrometer and JSM-6490LV scanning
electron microscopy (SEM). Exemplarily, one of the sample's SEM micrograph is
shown in Fig. 1. Also, AC and DC magnetization measurements were used to study
superconducting and magnetic properties of the samples. In all transport
measurements the standard four-point configuration with the low-frequency ac
technique was used. The resistance was measured with a current of $\sim$%
50$\mu$A. For the sample of the same (nominal) composition, sample-to-sample
fluctuations in magnitude of the critical temperature and the critical current
have been detected.

\textit{Experimental results}. (i) A representative temperature dependence of
the normalized resistivity R(T)/R(T = 39K) of the (bulk) samples with
MgB$_{2}$:La$_{0.67}$Sr$_{0.33}$MnO$_{3}$ weight ratio 1:1, 3:1 and 4:1, as
well as of the MgB$_{2}$ sample without manganite are shown in Fig. 2 (main
panel). According to the R(T) dependences, below the superconducting
transition temperature of MgB$_{2}$, samples 3:1 and 4:1 resistivity sharply
decreases and vanishes at critical temperature T$_{C}$ = 20 K. The resistivity
of sample 1:1 also decreases, but the composite of this composition still
remains in a resistive state down to the lowest temperature 4.2K. Sometimes a
hysteretic shift in R(T) behavior reflecting the (magnetic) pre-history
dependence of the transition temperature was also detected. Below T$_{C}$ the
results for the sample 4:1 basically reproduce those for the composition 3:1,
and below we concentrate on the results obtained for samples with weight ratio 3:1.

Figure 3 (main panel) shows the current-voltage (I -- V) characteristic of the
sample 3:1, taken at 4.2K. We observe a clear zero resistance supercurrent
branch, with a maximum value for I$_{C}$ of 2.5 mA. From the I - V
characteristics the critical current I$_{C}$ of the composite was determined
as the first deviation from a vertical line around zero bias. The second
deviation from linearity around 0.6 mV bias, with a maximum value of 8.9 mA,
is attributed to the critical current of the MgB$_{2}$ powder. In the current
interval 2.5 mA
$<$
I
$<$
8.9 mA the sample is in a resistive state and could be considered as a system
of small superconductive islands within a normal matrix. For comparison, in
Fig. 3 (inset) the I -- V characteristic of sample 1:1 at 4.2K is also shown.
This system remains in a resistive state, however, an excess current has been
clearly observed. Formally, the system is similar to arrays of resistively
shunted Josephson junctions.

The 3:1 composite magnetic susceptibility $\chi$ and magnetization M plotted
as temperature $\chi$(T) and field M(H) dependences are shown in Fig. 4 (a)
and Fig. 4 (b), respectively. As follows from the data for $\chi$(T) in Fig 4
(a), with decreasing temperature, a transition into ferromagnetic state occurs
with T$_{Curie}$ = 320 K and then, below 39K, a diamagnetic response is
developed. The measured magnetization hysteresis loop [see Fig.4 (b)] provides
the unambiguous evidence that our composite is a type-II superconductor with a
strong vortex pinning. At low temperature, with increasing an applied field, a
diamagnetic response is suppressed and above 3.3 kOe a ferromagnetic response
is again restored due to flux penetration and magnetization of the sample. 

(ii) Point-contact spectroscopy allows a direct measurement of the
superconducting gap. Figure 5 shows representative spectra of In, Ag, Nb --
sample 3:1 PCs measured at T = 4.2 K. We underline that the PCs' resistivity
varied by order, but multiple gaps structure in the quasiparticle density of
states of the composite, as well as the gap energy magnitudes were robust
features of superconducting composites and were reproduced in all PCs we
prepared. Conductance peaks corresponding to three superconducting gaps with
energies $\Delta_{1}(\pi)$ = 2.0 $\div$ 2.4 meV, $\Delta_{2}(\sigma)$ = 8.4
$\div$11.7 meV, and $\Delta_{tr}$ = 19.8 $\div$ 22.4 meV are clearly observed.
Here we will adopt abbreviation $\Delta_{1}(\pi)$ and $\Delta_{2}(\sigma)$ for
gaps which originate (most probably) from the $\Delta_{\pi}$ and
$\Delta_{\sigma}$ gaps of MgB$_{2}$, respectively, and $\Delta_{tr}$ which we
attribute (see discussion below) to the superconductivity of LSMO. It is worth
underlining here that magnitude of $\Delta_{tr}$ is the same as those earlier
detected in PCs of (La,Sr)MnO$_{3}$ and (La,Ca)MnO$_{3}$ with Pb or MgB$_{2}$
\cite{Kr06, Kr07}. In Fig. 6, the energy gap $\Delta_{tr}$ temperature
dependence is shown. From the BCS relation $\Delta$(0) = 1.76\textit{k}$_{B}%
$T$_{C}$, the $\Delta_{tr}$(0) gap would lead to a S with T$_{C}$ $\approx$
120 K. The classic BCS gap temperature behavior is shown in the figure too. As
is evident, the experimental data does not follow the BCS dependence.

Thus, principal features which arise from the data obtained in all PCs are (i)
a triple-gap structure of the PCs spectra; (ii) the\textit{ enhanced}
MgB$_{2}$ gaps. (Note , to avoid confusion, that within the accuracy of our
measurements, the magnitude of the $\Delta_{1}(\pi)$ gap still remains in
range of a bulk MgB$_{2}$ $\Delta_{\pi}$ gap \cite{XX}.) Also, (iii) proximity
induced quasiparticle gap $\Delta_{tr}$ is more than three times larger than
the conventional $\Delta_{\sigma}$ gap of MgB$_{2}$.

\textit{Discussion}. Here we will concentrate on temperature region T
$<$
T$_{C}$ and superconducting state of the 3:1 and 4:1 samples. As already
mentioned, an origin of the large quasiparticle $\Delta_{tr}$ gap and an
enhancement of the largest MgB$_{2}$ gap, $\Delta_{\sigma}$, cannot be
conventionally explained on the basis of the existing theoretical models of
proximity effect. Theoretically, the proximity effect in normal metal --
multi-band (MgB$_{2}$) hybrid structures has been considered by Brinkman
\textit{et al}., \cite{Br04}. A phenomenon was predicted, where the
superconductivity in a two-band superconductor is enhanced by the proximity to
a superconductor with a lower transition temperature (S'). The physics of this
effect is explained by the coupling between S' and the $\pi$-band: S'
superconductor enhances the superconductivity in the $\pi$-band, whereas the
$\pi-\sigma$ interband coupling causes an enhancement of the superconductivity
in the $\sigma$-band. However, for MgB:LSMO composite we found three
quasiparticle gaps, the largest of which, $\Delta_{tr}$, could not be
attributed to MgB$_{2}$. Also, a substantial increase in the magnitude of the
largest of MgB$_{2}$ gap, $\Delta_{\sigma}$, has been detected. Thus, a
question arises as to which of these gaps is proximity induced?

Before giving a qualitative explanation of the observations, we summarize the
physics of proximity effect at spin-active hmF/S interface which looks as
follows (see, e.g., \cite{Eschrig} and references therein). The conversion
process between the singlet and equal-spin triplet supercurrents is believed
to be governed by two important phenomena taking place in the half metal at
the interface: (i) a spin mixing and (ii) a spin-rotation symmetry breaking.
Spin mixing is the result of different scattering phase shifts that electrons
with opposite spin acquire when scattered (reflected or transmitted) from an
interface. It results from either a spin-polarization of the interface
potential, or differences in the wave-vector mismatches for spin up and spin
down particles at either side of the interface, or both. It is a robust and
ubiquitous feature for interfaces involving strongly spin-polarized
ferromagnets. 

Broken spin-rotation symmetry leads to spin-flip processes at the interfaces.
Its origin depends on microscopic magnetic state at the hmF/sS interface,
character of local magnetic moments coupling with itinerant electrons, etc.,
and even varies from sample-to-sample. However, the exact microscopic origin
of spin-flip processes at the interface is important only for the effective
interface scattering matrix \cite{Eschrig} and not for superconducting
phenomena, since Cooper pairs are of the size of the coherence length $\xi$
which is much larger than the atomic scale. 

Due to spin mixing at the interfaces, a spin triplet (S = 1, m = 0) amplitude,
$f_{\uparrow\downarrow}^{tr}(r)=(|\uparrow\downarrow>+|\downarrow\uparrow>)$,
is created and extends from the interface about a length $\xi_{F}$ = (D$_{F}%
$/2$\pi$H$_{exc}$)$^{1/2}$ into the F layer (D$_{F}$ denotes the diffusion
constant in the F). At the same time, triplet pairing correlations with equal
spin pairs (S = 1, m = +1 or m = 1) are also induced (due to spin-flip
processes) in the F layer. These components decay on the length scale $\xi
_{T}$ = (D$_{F}$/2$\pi$T)$^{1/2}$ which is much larger than $\xi_{F}$ because
in typical cases exchange field H$_{exc}$ is much larger than T$_{C}$. It is
worth emphasizing that it is only the m = 0 triplet component that is coupled
via the spin-active boundary condition to the equal-spin m = 1 pairing
amplitudes in the half-metal. The singlet component in the superconductor,
$f_{\uparrow\downarrow}^{s}(r)=(|\uparrow\downarrow>-|\downarrow\uparrow>)$,
being invariant under rotations around any quantization axis, is not directly
involved in the creation of the triplet m = 1 pairing amplitudes in the
half-metal. 

Note, that for odd-frequency \textit{s}-wave spin-triplet pairing \cite{Ef}
the static gap is zero. We observe nonzero gaps magnitudes, and thus this
scenario of proximity effect is not related to our case.

We now proceed to our system. Within the double-exchange interaction model for
hole-doped manganites (see, e.g., \cite{man}), the itinerant charge carriers
provide both the magnetic interaction between magnetic moments of the nearest
Mn$^{3+}$ - Mn$^{4+}$ ions and the system's electrical conductivity. Due to
the short mean free path, that is typically the distance of about a lattice
parameter, the charge carrier probes the magnetization on a very short length
scale. At the interfaces with the superconductor, the local moments are
expected to show a certain degree of disorder and it is reasonable to expect
that the interface ferromagnetic manganite -- \textit{s}-wave S is a
spin-active one. (Note also that local magnetic structure at the hmF/sS
interface can be easily affected by an external influence that leads to
hysteretic behavior of the equilibrium properties as we indeed observe in some
cases.) Following to Refs.\cite{Kr06, Kr07}, we suggest that, at low
temperature, non-coherent \textit{p}-wave even-frequency spin-triplet
superconducting condensate already exists in half-metallic manganites. Being
proximity coupled to the singlet superconductor, the m = 0 triplet component
in the hmF is coupled via boundary condition to the singlet pairing amplitude
in the S partner. At the same time, the spin-active boundary leads to coupling
of the m = 0 triplet component with an equal spin m = 1 pairing amplitude in
manganite. These couplings yield phase-coherency of both m = 0 and equal-spin
triplet Cooper pairs in the hmF with large quasiparticle gap $\Delta_{tr}$ (%
$>$
$\Delta_{\pi}$, $\Delta_{\sigma}$). As an inverse effect, being proximity
linked to \textit{s}-wave pairing amplitude, the m = 0 amplitude of the
triplet superconducting state enhances the quasiparticle gap(s) of a singlet superconductor.

In worthy to note here that in mean field BCS-Eliashberg theories, the
temperatures of electron-pairing, T$_{\Delta}$, and the long-range phase
coherency, T$_{\varphi}$, coincide and yield the critical temperature, T$_{C}%
$. This means that the global phase coherency and energy gap appear/vanish at
the same temperature, mainly due to the opening/disappearing of the gap with
temperature (i.e., the relation T$_{\varphi}$
$>$
T$_{\Delta}$ is a typical case). However, it has been shown \cite{Em95} (see
also \cite{Berg08}) that for systems with low conductivity and small
super-fluid density (bad metals), the temperature of the global phase
coherency T$_{\varphi}$ is reduced significantly and becomes comparable to or
even smaller than the pairing temperature T$_{\Delta}$. In this case, the
critical temperature T$_{C}$ is determined by the global phase coherency,
whereas the pair condensate could exist well above T$_{C}$. The experimental
results shown in Fig. 6 support the picture where the metastable
superconducting islands of nonzero order parameter are frozen, while the
long-range coherence associated with a bulk superconducting state is prevented
by marked fluctuations in the phase. (The loss of long-range phase stiffness,
rather than the vanishing of the pair condensate well above T$_{C}$, is at
present actively discussed in relation to several families of hole-doped
cuprates; see, e.g. \cite{Li10, Bi11} and references therein.)

The important consequence of the presence of the Cooper pair fluctuation above
T$_{C}$ is an appearance of the so-called pseudo-gap \cite{Em95, Lokt, N05},
i.e., decreasing of the one-electron density of state near the Fermi level. A
large pseudo-gap is indeed detected in numerous experiments on manganites
\cite{man} and it may be suggested that at least a portion of the observed
pseudo-gap is due to pairing without global phase coherency. Diamagnetic
response above T$_{C}$ provides additional arguments for survival of the pair
condensate well above T$_{C}$ in cuprates \cite{Li10} as well as in disordered
MgB$_{2}$ \cite{Ber08} and oxy-pnictides \cite{Pra11}. However, for manganites
this kind of response will be strongly suppressed by ferromagnetic order of
the localized moments and spin-triplet state of the condensate. 

\textit{Summary}. We produced and studied magnetic and superconducting
properties of MgB$_{2}$ -- (nano) La$_{0.67}$Sr$_{0.33}$MnO$_{3}$ composite.
Unconventional superconductivity of samples MgB:LSMO with 3:1 and 4:1 weight
ratio with T$_{C}$ = 20 K has been observed. Three distinct quasiparticle
energy gaps $\Delta_{1}(\pi)$, $\Delta_{2}(\sigma)$, and $\Delta_{tr}$ are
clearly revealed. Two of these gaps were identified as gaps originating from
the MgB$_{2}$. The third gap $\Delta_{tr}$, which is a few times larger than
the largest MgB$_{2}$ gap, could not be attributed to this \textit{s}-wave
superconductor. In our opinion, the results obtained testify in favor of the
induced long-range phase coherency of \textit{p}-wave even frequency spin
triplet superconductivity of half-metallic La$_{0.67}$Sr$_{0.33}$MnO$_{3}$.
Specifically, in a ferromagnetic state of La$_{0.67}$Sr$_{0.33}$MnO$_{3}$, a
condensate lost phase stiffness and is incapable to display long-range
supercurrent response, even though the gap amplitude is large. Nonetheless,
reduced local phase rigidity survives. Being proximity coupled to MgB$_{2}$,
the long-range coherency is restored. An inverse effect of the proximity
induced superconducting state with large gap is an enhancement of the
MgB$_{2}$ superconducting state. I.e., here we deal with some kind of "mutual"
proximity effect.

The authors acknowledge I. A. Danilenko for the LSMO nanoparticles
preparation. We also are grateful to M. Kupriyanov, Ya. Fominov for useful
discussions, and M. Belogolovskii for helpful comments on the manuscript.

\begin{center}
Figure Captions
\end{center}

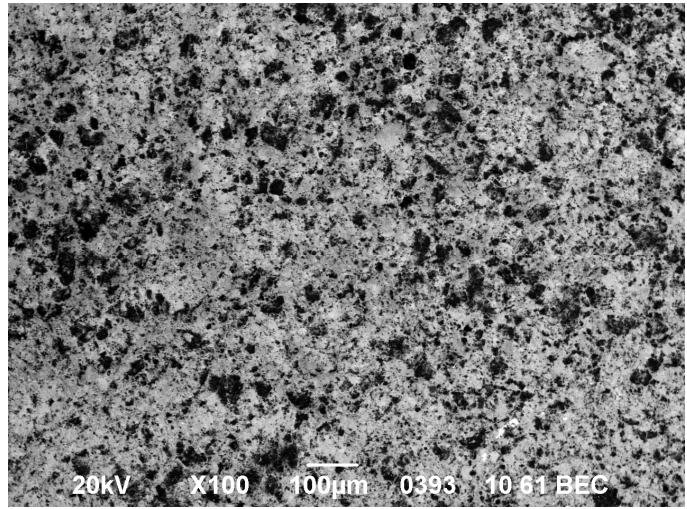
Fig. 1. SEM image of MgB$_{2}$:La$_{0.67}$Sr$_{0.33}$MnO$_{3}$ composite with
3:1 weight ratio (black -- MgB$_{2}$, light tone - La$_{0.67}$Sr$_{0.33}%
$MnO$_{3}$).

Fig. 2 Temperature dependence of the normalized resistivity R(T)/R(T = 39K) of
the samples with MgB$_{2}$:La$_{0.67}$Sr$_{0.33}$MnO$_{3}$ weight ratio 1:1,
3:1 and 4:1, as well as of pure (weight ratio 1:0) MgB$_{2}$. \textit{Inset}:
samples resistivity at T = 300K.

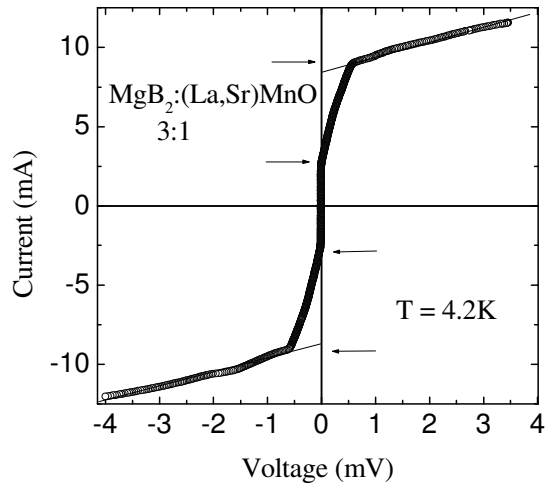
Fig. 3. (a) Current-voltage characteristic of MgB$_{2}$:La$_{0.67}$Sr$_{0.33}%
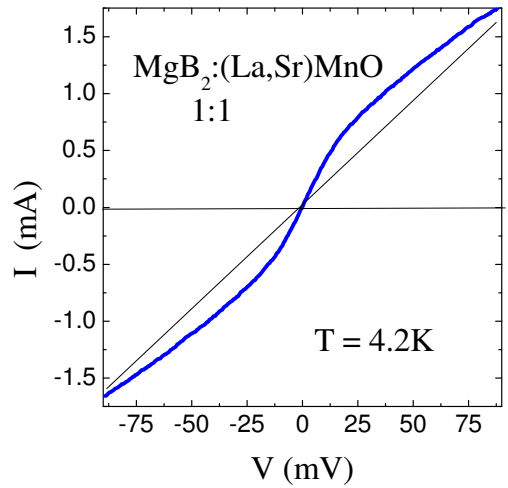
$MnO$_{3}$ (3:1) sample.\textit{ }(b)\textit{ }Current-voltage characteristic
of MgB$_{2}$:La$_{0.67}$Sr$_{0.33}$MnO$_{3}$ (1:1) sample.

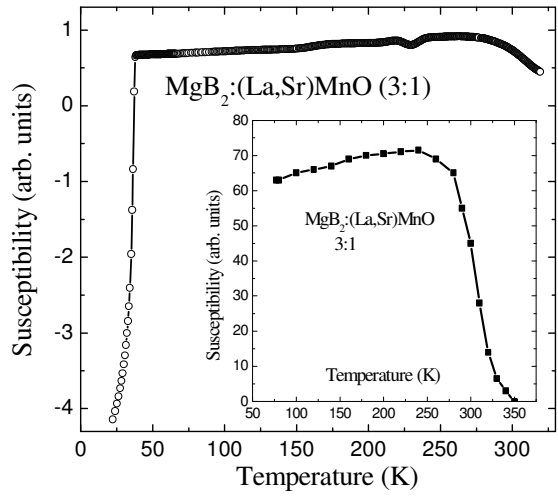
Fig. 4. (a) Temperature dependence of the magnetic susceptibility of MgB$_{2}%
$:La$_{0.67}$Sr$_{0.33}$MnO$_{3}$ (3:1) sample. \textit{Inset:} high
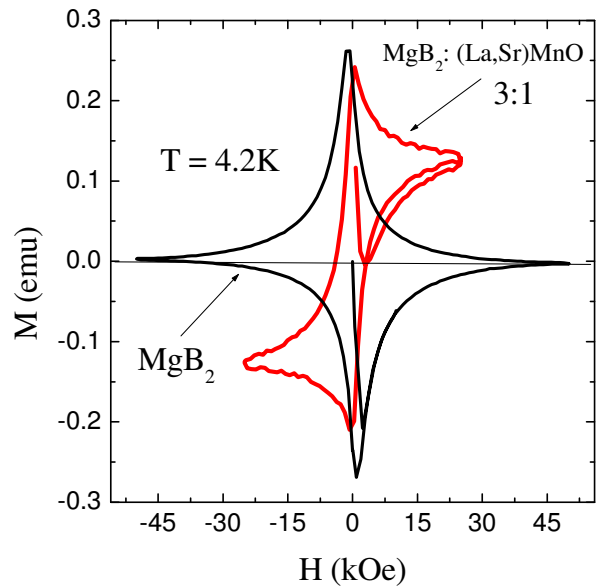
temperature region. (b) Magnetization M(H) curves of MgB$_{2}$:La$_{0.67}%
$Sr$_{0.33}$MnO$_{3}$ (3:1) sample and MgB$_{2}$.

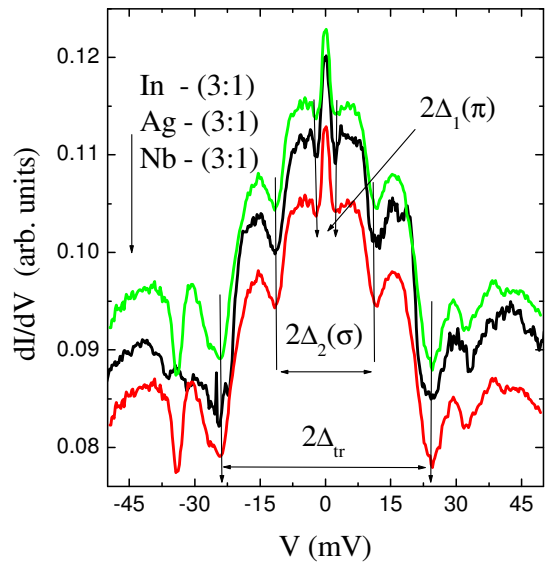
Fig. 5. The spectra of In -- sample 3:1, Ag -- sample 3:1 and of Nb -- sample
3:1 point contacts.

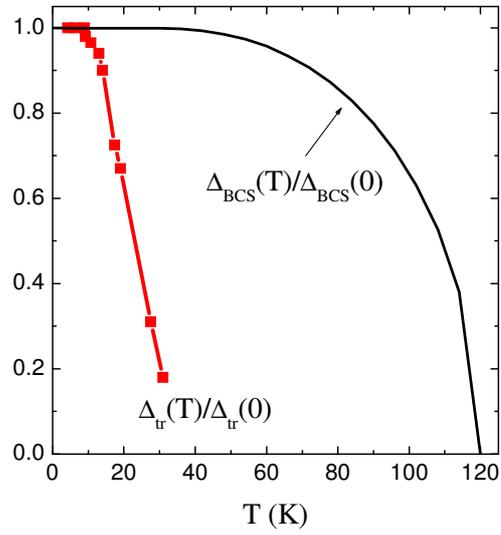
Fig. 6. Gap $\Delta_{tr}$(T) values (relative to zero temperature gap) of
MgB$_{2}$:La$_{0.67}$Sr$_{0.33}$MnO$_{3}$ (3:1) composite as function of
temperature and the BCS gap temperature dependence.

\end{document}